\documentclass[lettersize,journal]{IEEEtran}
\usepackage{amsmath,amssymb,amsfonts}
\usepackage{algorithmic}
\usepackage{algorithm}
\usepackage{array}
\usepackage[caption=false,font=footnotesize,labelfont=rm,textfont=rm]{subfig}
\usepackage{textcomp}
\usepackage{stfloats}
\usepackage{url}
\usepackage{verbatim}
\usepackage{graphicx}
\usepackage{cite}
\usepackage{xcolor}
\usepackage{subfig}
\usepackage{titlesec}
\usepackage{mathrsfs}
\usepackage{bm,bbold}
\usepackage{xcolor}
\begin{document}
	\title{{DRL-Based Antenna Position Optimization For MA-Assisted OTFS System Under Imperfect CSI}}
	
	\author{Maoyuan Wang, Qian Zhang, \IEEEmembership{Graduate Student Member,~IEEE}, Yufei Zhao, \IEEEmembership{Member, IEEE}, Xuejun Cheng, Zheng Dong, \IEEEmembership{Member, IEEE}, Deqiang Wang, \IEEEmembership{Member, IEEE}, and Yong Liang Guan, \IEEEmembership{Senior Member, IEEE}
		
		\thanks{Maoyuan Wang and Qian Zhang contributed equally to this work.}
		\thanks{Maoyuan Wang, Qian Zhang, Xuejun Cheng, Zheng Dong, and Deqiang Wang are with the School of Information Science and Engineering, Shandong University, Qingdao 266237, China (e-mail: \{maoyuanwang2024, qianzhang2021, chengxuejun\}@mail.sdu.edu.cn; \{zhengdong, wdq\_sdu\}@sdu.edu.cn).}
		\thanks{Yufei Zhao and Yong Liang Guan are with the School of Electrical and Electronics Engineering, Nanyang Technological University, Singapore (e-mail:\{yufei.zhao, eylguan\}@ntu.edu.sg).}
		\thanks{Corresponding author: Zheng Dong.}
	}
	
	\markboth{}
	{Shell \MakeLowercase{\textit{et al.}}: Bare Demo of IEEEtran.cls for IEEE Journals}
	\maketitle
	\begin{abstract}
		In this paper, we introduce movable antenna (MA) technology into orthogonal time frequency space (OTFS) systems to enable wavelength-level antenna position optimization under imperfect channel state information (CSI), thereby mitigating deep fading. To accurately acquire CSI, we develop a sparse Bayesian learning method with variational inference (SBLVI) method. Based on estimated CSI, we formulate an MA position optimization problem with the objective of maximizing channel gain. Due to the highly non-convex character of the problem, we further develop a deep reinforcement learning (DRL) strategy to intelligently optimize MA positions. Simulation results show that the proposed SBLVI method significantly improves channel estimation accuracy over benchmark methods, and MA position optimization based on estimated CSI achieves substantially higher channel gains than the fixed-position antenna (FPA), demonstrating the effectiveness of the proposed MA-assisted OTFS system.
	\end{abstract}
	
	
	\IEEEpeerreviewmaketitle
	\vspace{-0.3cm}
	
	\section{Introduction}
	The sixth-generation (6G) mobile network aims to provide global coverage, better intelligence level and improve data security, thereby enabling diverse emerging applications in high-mobility and even hostile environments~\cite{wang2022vision}. However, under such conditions, conventional modulation scheme such as orthogonal frequency-division multiplexing (OFDM), is susceptible to mobility-induced Doppler spread~\cite{wang2021pilot}. To address this challenge, orthogonal time-frequency space (OTFS) modulation has recently emerged as a promising technique to achieve robust and reliable communication in high-mobility scenarios~\cite{deng2025unifying}. Unlike OFDM, OTFS maps information symbols in the delay-Doppler~(DD) domain and spreads them over the entire time-frequency~(TF) domain, effectively mitigating Doppler effects in high-mobility environments. Furthermore, the authors in~\cite{liu2020uplink} explored OTFS-based massive multiple-input multiple-output (MIMO) for high-mobility communications. Nevertheless, the performance of such systems remains constrained by the fixed position of antennas.
	
	To explore the spatial variation of wireless channels with a limited number of antenna elements, the concept of movable antenna (MA) received much attention recently~\cite{zhu2023modeling, zhang2024efficient, xiu2025latency, zhu2023movable2, ma2024movable2, xiao2024channel}. By jointly optimizing antenna positions and beamformer, both communication and sensing performance can be significantly enhanced~\cite{ma2024multi3, zhu2023movable3, xiu2025movable3}. The authors in~\cite{zhu2023movable} proposed an MA enhanced multiple access design aiming to reduce the total transmission power of users, while their approach overlooks the challenges of dynamic environments characterized by mobility and time-varying channels. The authors in~\cite{xiu2026robust2} proposed an MA-aided cell-free integrated sensing and communication (ISAC) architecture to improve system performance under time-synchronization errors, while operating under the assumption of ideal channel state information (CSI). Existing studies have mainly considered designs under static environments with perfect CSI, and their applicability to dynamic environments and imperfect CSI scenarios remains limited. Fortunately, combining MA with OTFS technology can mitigate deep fading and thus enable high-quality communication in dynamic environments. In addition, optimal antenna position matching for MA requires accurate CSI to avoid the mismatch that may arise in practical applications.
	
	The antenna position optimization problem was restructured as a path planning problem in~\cite{mei2024movable}. A two-loop iterative algorithm for particle swarm optimization (PSO) was proposed to optimize antenna positions in~\cite{xiao2024multiuser}. The authors in~\cite{xie2024deep} used deep learning (DL) to optimize antenna positions. However, most traditional antenna positioning methods rely on static assumptions or pre-collected datasets, limiting their applications in dynamic scenarios. In contrast, we utilize deep reinforcement learning (DRL), which interacts with the environment and learns the positioning strategy through trial and error. Through adaptive learning, the DRL adapts to varying channel conditions and dynamically optimizes the MA position.
	
	This letter aims to utilize DRL to intelligently control the movement of the MA under imperfect CSI. To acquire channel path coefficients, we propose a sparse Bayesian learning with variational inference (SBLVI) method for OTFS system. Based on the estimation, we formulate an optimization problem that maximizes channel gain while jointly considering antenna position range, antenna movement distance per time step, and the channel between transmitter and receiver. Due to the highly non-convex form of the problem, we employ DRL to address the issue. Simulations demonstrate that MA controlled by the DRL method achieves higher channel gains than conventional fixed-position antenna (FPA), and the proposed SBLVI method obviously achieves lower normalized mean square error (NMSE) compared to linear minimum mean square error (LMMSE) and threshold-based embedded pilot (EP) methods.
	
	\section{System Model}
\begin{figure}
		\centering
		\includegraphics[width=0.72\linewidth]{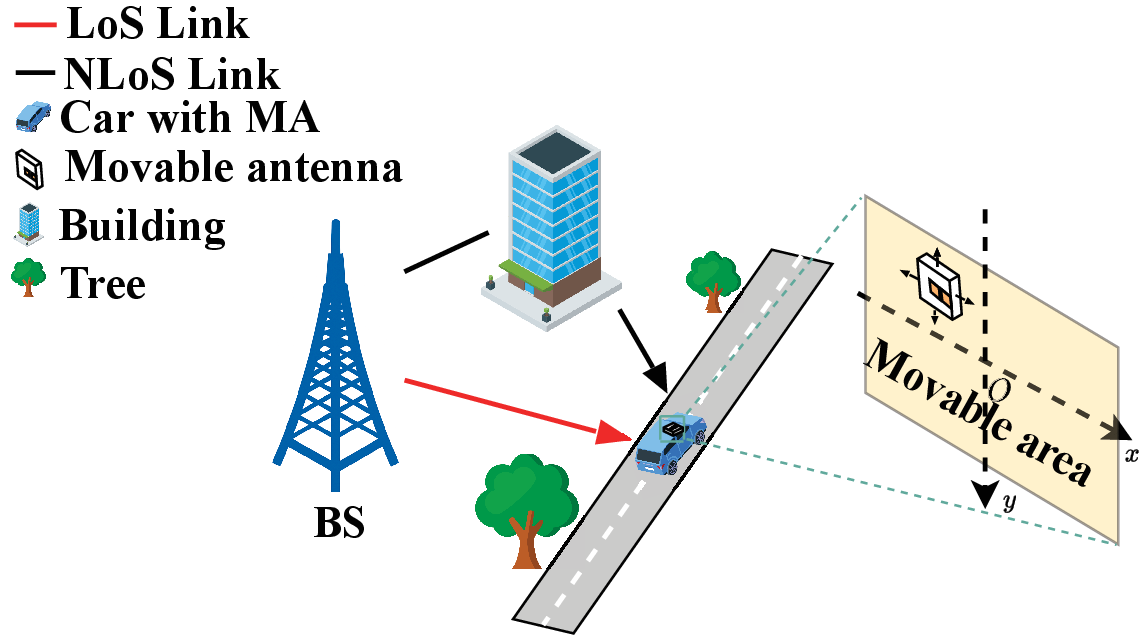}
		\caption{Model of a communication system assisted by movable antenna.}
		\label{system_figure}
		\vspace{-0.3cm}
	\end{figure}

	We consider a movable antenna-assisted OTFS downlink communication system, as shown in Fig.~\ref{system_figure}, where the MA for receiving is installed on the car moving at a speed~$v$. ‌The transmitter antenna at the base station (BS) remains stationary, while the MA operates within a two-dimensional square grid. The grid spans the range  $(-\lambda, \lambda)$  along both the $x$ and $y$ axes, where $\lambda$ denotes the wavelength. In this work, we consider multipath channels in dynamic environments. 
	
	Let $\mathbf{X}$, $\mathbf{Y} \in \mathbb{C}^{M \times N}$ be the transmitted and received two-dimensional delay-Doppler grids, respectively. Let $\mathbf{x}_{m} \in \mathbb{C}^{N \times 1}$ be column vectors containing the symbols in the $m$-th row of $\mathbf{X}$, where $m$ denotes the delay (row) indices in the two-dimensional grid, $\mathbb{C}$ denotes the set of all complex numbers. The total frame duration and bandwidth of the transmitted OTFS signal frame are $T_{f} = NT$ and $B = M\Delta f$, respectively, we consider $T\Delta f = 1$. Let $\mathbf{F}_{N}$ denotes the normalized $N$ point discrete Fourier transform (DFT) matrix and $\mathbf{F}_{N}^{\dagger}$ denotes the inverse discrete Fourier transform (IDFT) matrix. The time-frequency domain matrix ($\mathbf{X}_{\rm{tf}}$) is $\mathbf{X}_{\rm{tf}} = \mathbf{F}_M \cdot \mathbf{X} \cdot \mathbf{F}^{\dagger}_{N}$. Using waveform $g_{\rm tx}(t)$, the transmitted signal is given by
	\begin{align}
		\label{S outpue time}
		\mathbf{S} = \mathbf{G}_{\rm tx} \cdot (\mathbf{F}_{M}^{\dagger} \cdot \mathbf{X}_{\rm tf}) = \mathbf{G}_{\rm tx} \cdot (\mathbf{X} \cdot \mathbf{F}_{N}^{\dagger}),
	\end{align}
	where the diagonal matrix $\mathbf{G}_{\rm tx}$ has the samples of $g_{\rm tx}(t)$ as its entries, i.e., $\mathbf{G}_{\rm tx}$ = diag$[g_{\rm tx}(0),g_{\rm tx}(T/M),\ldots,g_{\rm tx}((M-1)T/M)] \in \mathbb{C}^{M \times M}$, where ${\rm diag}\{\mathbf{\cdot}\}$ denotes a diagonal matrix with the element in row $i$ and column $i$ equal to the $i$-th element of vector, and $(\cdot)^{\rm T}$ denotes transpose. Let $\mathbf{\tilde{X}}$ be the matrix containing the delay-time samples as $\mathbf{\tilde{X}}^{\rm T} = \mathbf{F}^{\dagger}_{N} \cdot \mathbf{X}^{\rm T}$.
	
	The time domain vector $\mathbf{s} \in \mathbb{C}^{NM \times 1}$, which is transmitted into the physical channel can be written as $\mathbf{s} = \text{vec}(\mathbf{G}_{\rm tx} \cdot \mathbf{\tilde{X}})$, where the samples are pulse shaped and transmitted as a continuous time signal $s(t)$, vec($\mathbf{A}$) denotes the column-wise vectorization of the matrix $\mathbf{A}$. At the receiver, the delay-time samples are obtained from waveform $\mathbf{r} \in \mathbb{C}^{NM \times 1}$ as
	\begin{align}
		\label{D receive signal}
		\mathbf{\tilde{Y}} = \text{vec}^{-1}_{N,M}((\mathbf{I}_{M} \otimes \mathbf{G}_{\rm rx}) \cdot \mathbf{r}),
	\end{align}
	where the diagonal matrix $\mathbf{G}_{\rm rx}$ has the samples of $g_{\rm rx}(t)$ as its entries, i.e., $\mathbf{G}_{\rm rx}$ = diag$[g_{\rm rx}(0),g_{\rm rx}(T/M), \ldots, g_{\rm rx}((M-1)T/M)] \in \mathbb{C}^{M \times M}$ is the pulse shaping filter at receiver, $\mathbf{r}$ denotes the sampled received time domain waveform as 
	\begin{align}
		\label{dis in_output relation}
		\mathbf{r}(q) = \sum_{l \in \mathcal{L}} g^{\rm s}(l,q) \mathbf{s}(q-l) + \mathbf{w}(q),
	\end{align}
	where $\mathbf{s}(q)$ means the $s(t)$ sampled at $t = qT/M$, $\mathbf{w}(q)$ is independent and identically distributed (i.i.d.) additive white Gaussian noise (AWGN) with variance $\delta^{2}_{n}$. $\otimes$ denotes the Kronecker product, $\mathbf{I}_{M}$ denotes an identical matrix of size $ M\times M$, and $\text{vec}^{-1}_{N,M}(\bf a)$ is the matrix formed by folding a vector $\mathbf{a}$ into a $N \times M$ matrix by filling it column wise.
	
	In our work, we assume $\mathbf{G}_{\rm tx} = \mathbf{G}_{\rm rx} = \mathbf{I}_{M}$ as in~\cite{liu2020uplink}. The output is given by
	\begin{align}
		\label{DD receive signal}
		\mathbf{Y} = \mathbf{F}^{\dagger}_{M} \cdot \mathbf{Y}_{\rm tf} \cdot \mathbf{F}_{N} = \mathbf{\tilde{Y}} \cdot \mathbf{F}_{N},
	\end{align}
	\begin{figure}
		\centering
		\includegraphics[width=1.0\linewidth]{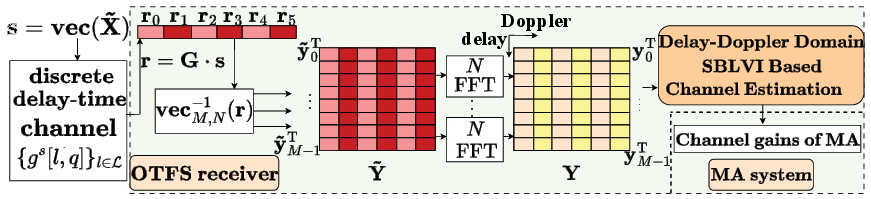}
		\caption{Discrete baseband model of the OTFS system for $N$ = 6, $M$ = 8 for receiver and MA system based on channel estimation.}
		\label{recei OTFS}
		\vspace{-0.3cm}
	\end{figure}
	moreover, we characterize the OTFS channel model as
	\begin{align}
		\label{dis delay time channel}
		g^{\rm s}(l,q) = \sum_{\ell \in \mathcal{L}^{\prime}} \left(\sum_{\kappa \in \mathcal{K}_{\ell}} \nu_{\ell}(\kappa) z^{\kappa(q-l)} \right) {\rm sinc}(l-\ell),
	\end{align}
	where $q$ and $l$ are obtained by sampling the received waveform at sampling intervals $t = qT/M$ for $0 \leq q \leq NM-1$ and discretizing delay variable $\tau = lT/ M$ for $0 \leq l \leq M-1$, respectively. Let $\ell_{i}$ and $\kappa_{i}$ denote the \emph{normalized delay shift} and \emph{normalized Doppler shift}, respectively, associated with the $i$-th path. The function ${\rm sinc}(x)$ is defined as ${\rm sinc}(x) = \frac{\sin(\pi x)}{\pi x}$ and $z = e^{\frac{j2\pi}{NM}}$. Let $\mathcal{L}^{\prime} = \{\ell_{i}\}$ be the set of distinct \emph{normalized delay shifts} among the $P$ paths in the delay-Doppler domain. For each $\ell \in \mathcal{L}^{\prime}$, define $\mathcal{K}_{\ell} = \{\kappa_{i} | \ell = \ell_{i} \}$ as the set of \emph{normalized Doppler shifts}. The $\nu_{\ell}(\kappa)$ can be denoted as
	\begin{align}
		\label{dd cof Vk}
		\nu_{\ell}(\kappa) = 
		\begin{cases}
			h_{i}, & \text{if } \ell = \ell_{i} \text{ and } \kappa = \kappa_{i} \\
			0, & \text{otherwise},
		\end{cases}
	\end{align} 
    where $h_{i}$ is the path complex coefficient. We assume that channel delay shifts can be modeled as integer delay shifts, thus \eqref{dis delay time channel} can be reduced to $g^{\rm s}(l,q) = \sum_{\kappa \in \mathcal{K}_{l}} \nu_{l}(\kappa) z^{\kappa(q-l)}$. Based on the $h_{i}$, we can evaluate the channel gain. Specifically, we use a path-response matrix (PRM) $\mathbf{\Sigma} \in \mathbb{C}^{P \times P}$ to represent the response from the transmit position to receive position\cite{zhu2023modeling}. It can be written as diagonal PRM $\mathbf{\Sigma} = {\rm diag}\{h_{1},h_{2},...,h_{P}\}$. We use a single antenna at the receiver, thus the difference of the signal propagation distance between position $(x,y)$ and reference point $(0,0)$ in the receiver is given by $\rho_{i}(x,y) = x \cos\zeta_{i} \sin\eta_{i} + y \sin\zeta_{i}$ for the $i$-th path. The field-response vector can be written as\footnote{We consider a single receive MA to explore the impact of DRL-based position optimization without the additional effects of multi antennas combining, and our current model can be extended to the multi antenna systems. If the receiver has $A$ MAs, which positions are $(x_{e}, y_{e})$, where $e=1,2,\ldots,A$, for the $e$-th antenna, the field-response vector is $\mathbf{f}(x_{e}, y_{e})=\left[ \exp(j \frac{2\pi}{\lambda}\rho_{1}(x_{e}, y_{e})),\ldots, \exp(j \frac{2\pi}{\lambda}\rho_{P}(x_{e}, y_{e})) \right]^{\rm T}$, and we have $\mathbf{F} =[\mathbf{f}(x_{1}, y_{1}), \ldots, \mathbf{f}(x_{A}, y_{A})]$, the channel gain is $\left\|\mathbf{U}(x,y)\right\|_{2}^{2}=\left\|\mathbf{F}^{\rm{H}}\mathbf{\Sigma} \mathbf{1}_{P}\right\|_{2}^{2}$.} 
	\begin{equation}
		\label{f(r)}
		\mathbf{f}(x,y) = \Big[\exp\big({j \frac{2\pi}{\lambda}\rho_{1}(x,y)}\big),...,\exp\big({j \frac{2\pi}{\lambda}\rho_{P}(x,y)}\big)\Big]^{\rm T}.
	\end{equation}

	We formulate the channel gain between the transmit and the receive antenna as $|U(x,y)|^{2} = |\mathbf{f}(x,y)^{\rm H} \mathbf{\Sigma} \mathbf{1}_{P}|^{2}$.
	
	\section{SBLVI Channel Estimation Based on OTFS System}
	After transmission through the channel, the received $\mathbf{Y}$ is obtained by applying the transformations in \eqref{D receive signal} and \eqref{DD receive signal}, and we can denote $\mathbf{y}$ as $\mathbf{y} = \text{vec} (\mathbf{Y})$. We can get the estimated angle of arrival (AOA) by fixing the transmitter and moving the receiver antenna \cite{ma2023compressed}, thus the angle information of the $P$ paths can be obtained. Furthermore, a key problem is how to obtain more accurate channel coefficient. To address this challenge, we use the SBLVI algorithm. The initial estimation of the delay and Doppler indices as 
	\begin{equation}
		\label{choose P largest}
		\{\mathbf{l}^{0},\mathbf{k}^{0}\} = \text{arg}\,\text{max}_{l,k}|\mathbf{Y}[l,k]|,
	\end{equation}
	where $\mathbf{l}^{0} = [l^{0}_{1}, l^{0}_{2}, \ldots, l^{0}_{P}]^{\rm T}$ and $\mathbf{k}^{0} = [k^{0}_{1}, k^{0}_{2}, \ldots, k^{0}_{P}]^{\rm T}$ denote the delay index and  Doppler index for the path at the initial, respectively. Then, let $\mu_{l,i} = l^{0}_{i} - L_{p}$ and $\mu_{k,i} = k^{0}_{i} - K_{p}$, where $L_{p}$ denotes number of rows of pilot, $K_{p}$ denotes number of columns of pilot, then we can get vectorization of $\mu_{l,i}$ and $\mu_{k,i}$, denoted as $\mathbf{l}_{i}$ and $\mathbf{k}_{i}$, where $\mathbf{l}_{i} = [l_{1}, l_{2}, \ldots, l_{P}]^{\rm T}$ and $\mathbf{k}_{i} = [k_{1}, k_{2}, \ldots, k_{P}]^{\rm T}$. We represent $f_0$ as number of iterations, $\epsilon$ denotes convergence metric.
	
	We denote $l_{i,p} = l_{i} + L_{p}$, $k_{i,p} = k_{i} + K_{p}$ as the delay index and Doppler index for $i$-th path, and construct the fractional delay basis functions for each path, for delay dimension as
	\begin{align}
		\label{Dirichlet interpolation core - corresponding delay}
		\omega_{\tau}[m,i] &= \frac{1}{M} \frac{\text{sin}(\pi(m - l_{i,p}))}{\text{sin}(\pi(m - l_{i,p})/M)}\nonumber\\
		& \qquad \times \exp\bigg(-{\frac{j\pi(M-1)(m - l_{i,p})}{M}}\bigg),
	\end{align}
	where $m = 1, \ldots, M$, which is the discrete delay indices, and $M$ denotes the number of discrete delay grid points. Similarly, for Doppler dimension, its basis functions as
	\begin{align}
		\omega_{\nu}[n,i] &= \frac{1}{N} \frac{\sin(\pi(n - k_{i,p}))}{\sin(\pi(n - k_{i,p})/N)} \nonumber\\
		&\qquad \times \exp\bigg(-\frac{j\pi(N-1)(n - k_{i,p})}{N}\bigg),\label{Dirichlet interpolation core - corresponding doppler}
	\end{align}
	where $n = 1, \ldots, N$ represents the Doppler indices, and $N$ denotes the number of Doppler grid points. $l_{i,p}$ and $k_{i,p}$ can be fractional, and we have 
	\begin{align}
		\label{partial derivative with respect to delay}
		\frac{\partial \boldsymbol{\phi}_{\tau}[m]}{\partial l_{i,p}} &= -\frac{1}{M} \sum_{u=0}^{M-1} j \frac{2\pi u}{M} \nonumber\\
		&\qquad \times \exp\bigg(\frac{j2\pi(m - l_{i,p})u}{M}\bigg),
	\end{align}
	where $u$ represents the summation index running from $0$ to $M-1$, and $\partial(\cdot)$ denotes the partial differential of a function. Then, we have
	\begin{align}
		\label{partial derivative with respect to doppler}
		\frac{\partial \boldsymbol{\phi}_{\nu} [n]}{\partial k_{i,p}} &= \frac{1}{N} \sum_{\varsigma=0}^{N-1} j\frac{2\pi\varsigma}{N}  \nonumber\\
		&\qquad \times \exp\bigg(\frac{-j2\pi(n-k_{i,p})\varsigma}{N}\bigg),
	\end{align}
	where $\frac{\partial \boldsymbol{\phi}_{\tau}[m]}{\partial l_{i,p}}$ and $\frac{\partial \boldsymbol{\phi}_{\nu} [n]}{\partial k_{i,p}}$ denote the partial derivative of the $m$-th and $n$-th sample of the delay and Doppler basis function for a fixed path, and $\varsigma$ represents the summation index running from $0$ to $N-1$. Then, $\mathbf{\Phi} = \text{vec}(\boldsymbol{\omega}_{\tau}(:,p) \mathbf{\boldsymbol\omega}^{\rm T}_{\nu}(:,p))$ constructs a two-dimensional basis function, $\boldsymbol{\Phi}_{\nu}(:,i) = \boldsymbol{\omega}_{\tau}(:,i) \otimes \frac{\partial \boldsymbol{\phi}_{\nu}}{\partial k_{i,p}}$ and $\boldsymbol {\Phi}_{\tau}(:,i) = \frac{\partial \boldsymbol{\phi}_{\tau}}{\partial l_{i,p}} \otimes \mathbf{\boldsymbol\omega}_{\nu}(:,i)$.
	
	In $f_0$-th iteration, updating posterior mean and covariance of channel coefficients using variational Bayes as $\mathbf{H}_{h} = \boldsymbol{\Phi}^{\rm H} \boldsymbol{\Phi} + \boldsymbol{\Phi}^{\rm H}_{\nu} \boldsymbol{\Phi}_{\nu} \odot \boldsymbol{\Sigma}_{k} + \boldsymbol{\Phi}^{\rm H}_{\tau} \boldsymbol{\Phi}_{\tau} \odot \boldsymbol{\Sigma}_{l}$, where $\boldsymbol{\Sigma}_{k}$ and $\boldsymbol{\Sigma}_{l}$ initial value is $\mathbf{I}_{P}$, $\boldsymbol{\Phi}^{\rm H}$ denotes the conjugate transpose of $\boldsymbol{\Phi}$. We have $\boldsymbol{\Sigma}_{h} = \left(\frac{c}{d}\mathbf{H}_{h} + {\rm diag}\left(\mathbf{a}\oslash\mathbf{b}\right)\right)^{-1}$ and $\boldsymbol{\mu}_{h} = \boldsymbol{\Sigma}_{h} \frac{c}{d} {\boldsymbol \Phi}^{\rm H} \mathbf{y}$, where the initial value of $\mathbf{a}$ and $\mathbf{b}$ is $\mathbf{1}_{P}$, it denotes an $P$-dimensional vector with all the elements equal to 1. The initial value of c and d is $10^{-6}$, $10^{-3}$, respectively. In addition, $\oslash$ represents element-wise division of a matrix or vector. $\boldsymbol{\mu}_{h} \in \mathbb{C}^{P \times 1}$ denotes posterior mean of channel coefficients, and $\boldsymbol{\Sigma}_{h}$ represents the posterior covariance of channel coefficients, such that $a_{i} \leftarrow a_{i} + 1$ and $\mathbf b \leftarrow \mathbf b + |{\boldsymbol\mu}_{h}|^{2} + \boldsymbol{\Sigma}_{h,i}$, in which $\boldsymbol{\Sigma}_{h,i}$ $\triangleq$ ${\rm diag}(\boldsymbol{\Sigma}_{h})$. Moreover, $c$ is denoted as $c \leftarrow c + MN$, and $d$ is written as $d \leftarrow d + \|\mathbf{y}\|^{2} - 2\Re(\mathbf{y}^{\rm H} \boldsymbol{\Phi} \boldsymbol{\mu}_{h})+ \Re(\boldsymbol{\mu}_{h}^{\rm H} \mathbf{H}_h \boldsymbol{\mu}_{h}) + tr(\mathbf{H}_{h} \odot \boldsymbol{\Sigma}_{h})$, where $\|\cdot\|$ represents the norm, $\Re(\cdot)$ denotes the real part, $\odot$ denotes Hadamard product. We can update $\boldsymbol{\Sigma}_{k}$ and $\boldsymbol{\Sigma}_{l}$ as
	\begin{align}
		\label{Sigma k and Sigma l}
		\boldsymbol{\Sigma}_k = \left(\frac{c}{d} \Re \big((\boldsymbol{\Sigma}_{h} + \boldsymbol{\mu}_{h} \boldsymbol{\mu}_{h}^{\rm H}) \odot (\boldsymbol{\Phi}^{\rm H}_{\nu} \boldsymbol{\Phi}_{\nu})\big)  \right)^{-1},\\
		\boldsymbol{\Sigma}_{l} = \left(\frac{c}{d} \Re \big((\boldsymbol{\Sigma}_{h} + \boldsymbol{\mu}_{h} \boldsymbol{\mu}_{h}^{\rm H}) \odot (\boldsymbol{\Phi}^{\rm H}_{\tau} \boldsymbol{\Phi}_{\tau})\big)  \right)^{-1},
	\end{align}
	updating delay indices $\hat{\mathbf{l}}$, we can get $\mathbf{h} \leftarrow \boldsymbol{\mu}_{h}$, $\mathbf{k}_{i} \leftarrow \hat{\mathbf{k}}$, $\mathbf{l}_{i} \leftarrow \hat{\mathbf{l}}$. So, we have channel matrix as $\mathbf{H}_{DD} = \boldsymbol{\omega}^{\rm T}_{\tau} \text{diag}(\mathbf{h}) \boldsymbol{\omega}_{\nu}$, where
	\begin{align}
		\label{omega tau and omega nu}
		\boldsymbol{\omega}_{\tau}& = \frac{1}{M} \exp\bigg(j \frac{(M-1)\pi (m-(l_{i} + L_{p}))}{M}\bigg) \nonumber\\
		&\quad \times \frac{\sin(\pi(m-(l_{i} + L_{p})))}{\sin(\pi(m-(l_{i} + L_{p}))/M)},\\
		\boldsymbol{\omega}_{\nu} &= \frac{1}{N} \exp\bigg(-j \frac{(N-1)\pi (n-(k_{i} + K_{p}))}{N}\bigg) \nonumber\\
		&\quad \times\frac{\sin(\pi(n-(k_{i} + K_{p})))}{\sin(\pi(n-(k_{i} + K_{p}))/N)},
	\end{align}
	the convergence check is performed as $\epsilon = \frac{\| \mathbf{H}^{(f_0+1)}_{DD} - \mathbf{H}^{(f_0)}_{DD}\|}{\|   \mathbf{H}^{(f_0)}_{DD}\|}$. We can get the complex coefficient of $i$-th path, denoted as $\mathbf{h}_{\rm max} = [h_1, h_2,\ldots, h_P]$. $\textbf{Algorithm 1}$ shows the corresponding operation, and the system diagram is shown in Fig. \ref{recei OTFS}.
	\begin{algorithm}[tbp]
		\label{SBLVI1}
		\caption{SBLVI channel estimation}
		\renewcommand{\algorithmicrequire}{\textbf{Input:}}
		\renewcommand{\algorithmicensure}{\textbf{Output:}}
		\begin{algorithmic}[1]
			\REQUIRE $\mathbf{Y}$, $M$, $N$, $P$, $L_{p}$, $K_{p}$, $F$ = 200, $\text{Epsilon} = 10^{-6}$
			\ENSURE $\mathbf{H}_{DD}$, $\mathbf{h}_{\rm max}$
			\STATE initialization\;
			\WHILE{$f_0<F \,\text{and } \epsilon > \text{Epsilon}$}
			\FOR{$i$ = 1:$P$}
			\STATE $\boldsymbol{\Phi} = \text{vec}(\boldsymbol{\omega}_{\tau}(:,i) \boldsymbol{\omega}^{\rm T}_{\nu}(:,i))$
			\STATE $\boldsymbol{\Phi}_{\nu}(:,i) = \boldsymbol{\omega}_{\tau}(:,i) \otimes \frac{\partial \boldsymbol{\phi}_{\nu}}{\partial k_{i,p}}$
			\STATE $\boldsymbol { \Phi}_{\tau}(:,i) = \frac{\partial \boldsymbol{\phi}_{\tau}}{\partial l_{i,p}} \otimes \boldsymbol{\omega}_{\nu}(:,i)$
			\ENDFOR
			\STATE $\mathbf{H}^{(f_0)}_{h} = \mathbf{\Phi}^{\rm H} \mathbf{\Phi} + \mathbf{\Phi}^{\rm H}_{\nu} \mathbf{\Phi}_{\nu} \odot \boldsymbol{\Sigma}^{(f_0-1)}_{k} + \mathbf{\Phi}^{\rm H}_{\tau} \mathbf{\Phi}_{\tau} \odot \boldsymbol{\Sigma}^{(f_0-1)}_{l}$
			\STATE $\mathbf{h}^{(f_0)} = \boldsymbol{\mu}^{(f_0)}_{h}$
			\STATE $\mathbf{k}^{(f_0)}_{i} = \mathbf{\hat{k}}^{(f_0)}$
			\STATE $\mathbf{l}^{(f_0)}_{i} = \mathbf{\hat{l}}^{(f_0)}$
			\STATE $\mathbf{H}^{(f_0)}_{DD} = \boldsymbol{\omega}^{T}_{\tau} {\rm diag}(\mathbf{h}^{(f_0)}) \boldsymbol{\omega}_{\nu}$
			\ENDWHILE		
		\end{algorithmic}	
	\end{algorithm}

	\section{MA Position Optimization Based on DRL}
	\subsection{Problem Formulation}
	At time slot $t$, there are $P_t$ paths, and $U(x,y,t)$ represents the channel $U(x,y)$ at time slot t. Our goal is to optimize antenna position for high channel gain, given by 
	\setcounter{equation}{18}
	\begin{subequations}
		\begin{align}
			\max_{x,y} \quad & |U(x,y,t)|^2 \\
			\text{s.t.} \quad & -\lambda \leq x_{t} \leq \lambda, \forall t \in \mathbb{Q}^{+}, \\
			& -\lambda \leq y_{t} \leq \lambda, \forall t \in \mathbb{Q}^{+}, \\
			& \|[x_{t+1}, y_{t+1}] - [x_t, y_t]\|_\infty \leq \Delta_{\text{max}}, \forall t \in \mathbb{Q}^{+}, \\
			& U(x,y,t) = \sum_{i=1}^{P_t} h_i(t) \times \exp\bigg(\frac{j2\pi}{\lambda}x_{t}\cos\zeta_{i,t}\sin\eta_{i,t}\bigg) \nonumber\\
			& \qquad \qquad \quad \times \exp\bigg(\frac{j2\pi}{\lambda}y_{t}\sin\zeta_{i,t}\bigg).
		\end{align}
	\end{subequations}

	Constraints (19b) and (19c) represent the range of MA is related to the wavelength $\lambda$. Constraint (19d) means that the range of movement at adjacent time slots cannot exceed the maximum movement distance $\Delta_{\rm max}$. Constraint (19e) indicates that the channel at different positions $(x_{t},y_{t})$ at different time slots $t$. However, the problem is highly non-convex, which is very difficult to solve. To address this problem, we utilize a Deep Q-Network (DQN) combined with SBLVI channel estimation method to optimize antenna position. 
	
	\subsection{DQN Algorithm for MA under Estimated Channel}
	We define the state space $\mathcal{S}$, action space $\mathcal{A}$, transition function $\mathcal{P}$, reward function $\mathcal{R}$, discount factor $\gamma$, and the time horizon $\mathcal{T}$ divided into a set of discrete time steps $\{1,2,...,T\}$.
	
	$\bullet \mathsf{~State~Space}~\mathcal{S}$: At time step $t$, each state $\psi_{t} \in \mathcal{S}$ is an eight-dimensional vector, $x_{t}$ and $y_{t}$ denote the MA position. $\mathbf{Z}_{t}$ denotes the average gain of the surrounding eight points that were visited, $\mathbf{V}_{t}$ denotes visit frequency of positions. $\varpi_{t}$ denotes the distance between the current point and the center point, $\varrho_{t}$ denotes the current step in an episode divided by the maximum number of steps, and $\chi_{t} = 1 - \varrho_{t}$. The state can be expressed as $\psi_{t} = \{ x_{t}, y_{t}, |U(x,y,t)|^{2}, \mathbf{Z}_{t}, \mathbf{V}_{t}, \varpi_{t}, \varrho_{t}, \chi_{t}, \forall t \in \mathbb{Q}^{+}\}$.
	
	$\bullet \mathsf{~Action~space}~\mathcal{A}$: At time step $t$, each action $\Lambda_{t} \in \mathcal{A}$ represents the antenna movement decision, it can be expressed as $\Lambda_{t} = \{\Lambda_{1}, \Lambda_{2}, \Lambda_{3}, \Lambda_{4},\Lambda_{5}\}$, from $\Lambda_{1}$ to $\Lambda_{5}$ they represent move up, down, left, right and stay still, respectively.
	
	$\bullet \mathsf{~Transition ~ function}~\mathcal{P}$:  $\mathcal{P}(\psi_{t+1}|\psi_{t}, \Lambda_{t})$ defines the environment $\psi_{t}$ changes over time due to the actions $\Lambda_{t}$.
	
	$\bullet \mathsf{~Reward ~ function}~\mathcal{R}$: The reward function is $\mathcal{R}_{t}(\psi_{t}|\Lambda_{t})$, the reward $\Gamma_{t}$ is determined as the channel gains $|U(x,y,t)|^{2}$.
	
	Based on $\psi_{t}$, the agent takes an action, gets a reward, and the environment moves to next state. Then save $\psi_{t}$, $\Lambda_{t}$, $\Gamma_{t}$, $\psi_{t+1}$ into experience replay buffer $\mathcal{D}$. The loss function as
	\begin{equation}
		\label{loss}
		\mathcal{L} (\theta)= \mathbb{E}_{\mathcal{D}}~\left(\Gamma + \gamma \max_{\Lambda'} Q(\psi',\Lambda';\theta^{-}) - Q(\psi,\Lambda;\theta)\right)^2,
	\end{equation}
	where $\mathbb{E}$ denotes the expectation, and we can denote $\iota$ as
	\begin{equation}
		\label{Q_{target}}
		\iota = \Gamma + \gamma \max_{\Lambda'} Q(\psi',\Lambda';\theta^{-}),
	\end{equation}
	where $\psi'$ represents next state, $\Xi_{t}$ denotes the episode termination flag, $\max Q(\psi',\Lambda';\theta^{-})$ means that maximum Q-value over all possible actions, $\theta^{-}$ is the parameter of the target network, $Q(\psi,\Lambda;\theta)$ denotes the Q-value for taking action $\Lambda$ in $\psi$, as shown in $\textbf{Algorithm 2}$.
	\begin{algorithm}[tbp]
		\label{DQN}
		\caption{DQN algorithm for movable antenna}
		\begin{algorithmic}[1]
			\STATE initialize replay memory $\mathcal{D}$ and random weight $\theta$
			\STATE initialize target action-value function $\hat{Q}$ with $\theta^{-} \leftarrow \theta$
			\FOR{episode = $1,2...$}
			\STATE Initialize random MA position $(x,y)$
			\FOR{ $t = 1,2,\ldots,T$}
			\STATE With probability $\epsilon$ select random action $\Lambda_{t}$
			\STATE Otherwise select $\Lambda_{t} = {\rm argmax}_{\Lambda} Q(\psi_{t},\Lambda;\theta)$
			\STATE Execute action $\Lambda_{t}$, move antenna to new position
			\STATE Estimate channel gain at new position
			\STATE Calculate reward $\Gamma_{t}$ and observe next state $\psi_{t+1}$
			\STATE Store transition $(\psi_{t}, \Lambda_{t}, \Gamma_{t}, \psi_{t+1}, \Xi_{t})$ in $\mathcal{D}$
			\STATE minibatch of $(\psi_{o}, \Lambda_{o}, \Gamma_{o}, \psi_{o+1}, \Xi_{o})$ \text{from} $\mathcal{D}$ 
			\STATE $\iota_{o} = 
			\begin{cases}
				\Gamma_{o} & \text{if terminal} \\
				\Gamma_o + \gamma \max\limits_{\Lambda'} \hat{Q}(\psi_{o+1}, \Lambda'; \theta^-) & \text{otherwise}
			\end{cases}$
			\STATE gradient descent step on $(\iota_{o} - Q(\psi_{o},\Lambda_{o};\theta))^{2}$ with respect to $\theta$, reset $\theta^{-} \leftarrow \theta$ 
			\ENDFOR 
			\ENDFOR
		\end{algorithmic}
	\end{algorithm}
	
	\section{Simulation Results and Discussion}
	
	In this section, we evaluate the performance of the proposed design by computer simulations. We assume that the channel remains quasistatic within each coherence block. The MA can be placed within a 101×101 grid. We set $M = N = 64$, the learning rate is denoted as $\alpha$. We place the FPA at the grid center. The discount factor $\gamma=$ 0.99, the initial exploration rate $\epsilon_{\rm start}=$ 1.0, the exploration rate decay factor $\epsilon_{\rm decay}=$ 0.997, $\Delta_{\rm max}=2\lambda/101$. Mini batch size $D = 128$, the maximum number of steps per episode $T = 100$, a total of 1000 episodes.
	
\begin{figure}[t]
		\centering
		\subfloat[The car speed $v=$ 70 km/h]{\includegraphics[width=0.495\linewidth]{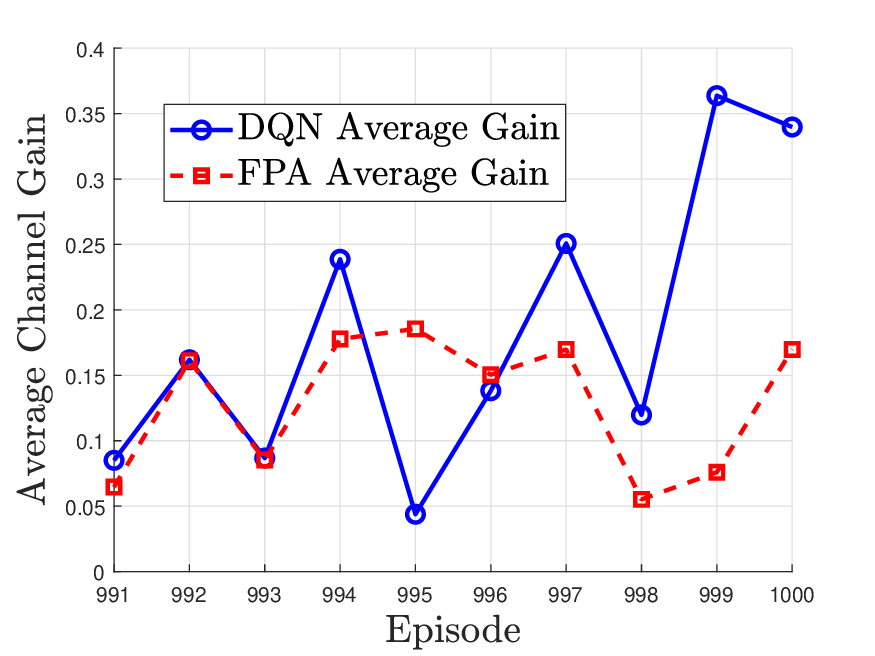}%
			\label{70kmph}}
		\hfil
		\subfloat[The car speed $v=$ 100 km/h]{\includegraphics[width=0.495\linewidth]{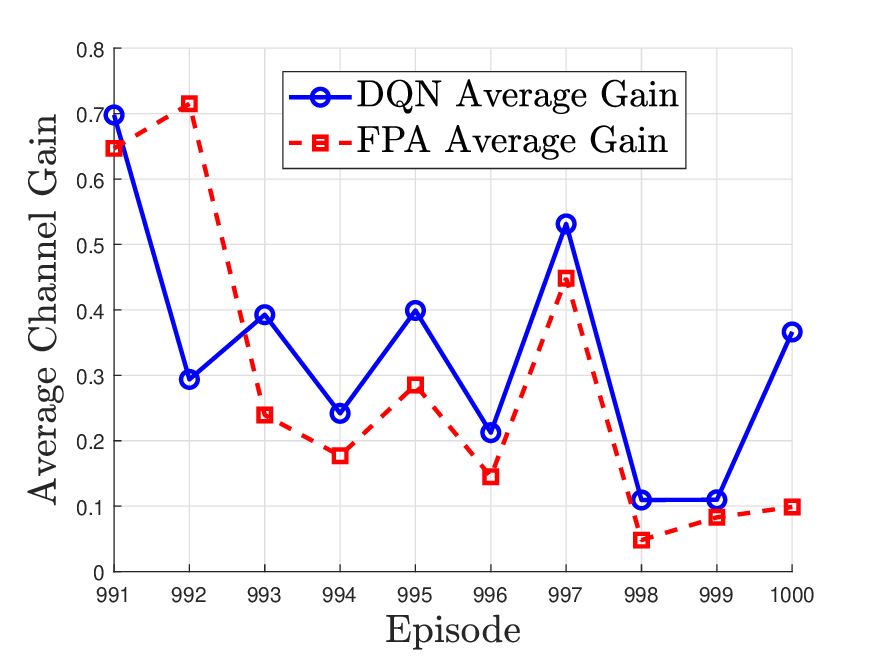}%
			\label{100kmph}}
		\hfil
		\caption{Comparison of the average channel gain between MA and FPA in different dynamic channel environments for $\alpha = 10^{-7}$.}
		\label{MAvsFPA}
		\vspace{-0.3cm}
	\end{figure}

	We compare the performance of MA and FPA after convergence in Fig.~\ref{MAvsFPA}. The average gain is to take the average of all step gains in each episode. In the first dynamic environment, the car speed is $v =$ 70 km/h, the DRL-based MA achieves a higher average gain than the FPA in 80\% of the episodes. In the second environments, MA also identifies favorable positions. In this case, the MA positions determined by the DRL-based approach outperform FPA in 90\% of episodes.
	
\begin{figure}[t]
		\centering
		\includegraphics[width=0.75\linewidth]{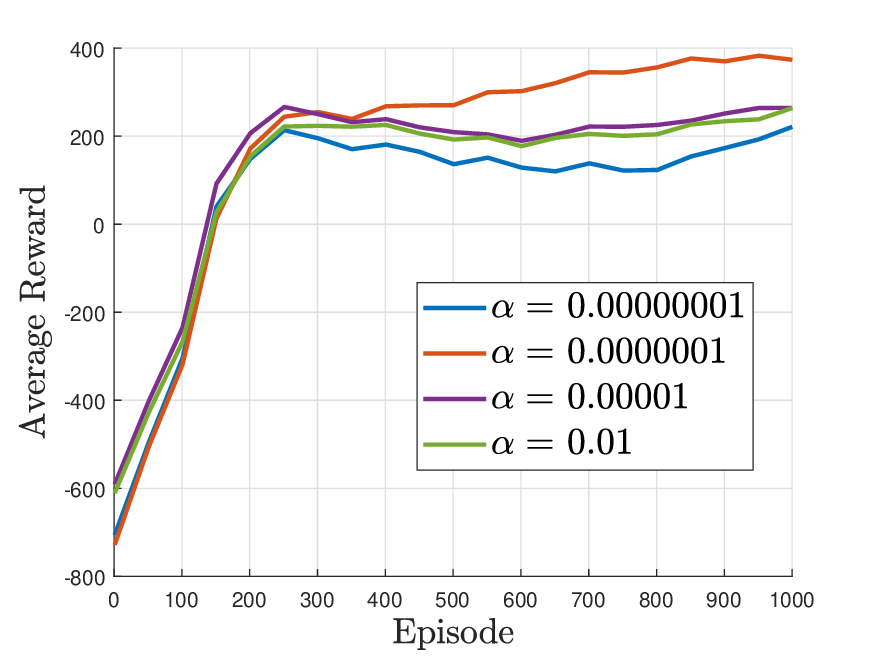}
		\caption{Convergence property with various learning rates.}
		\label{lrcompcur}
		\vspace{-0.3cm}
	\end{figure}

	We depict different learning rates in Fig.~\ref{lrcompcur}. DRL with a learning rate of $10^{-7}$ takes longer to converge than $10^{-5}$, but it still achieves the best performance. If the learning rate is too small as $\alpha= 10^{-8}$, the system takes longer to converge, resulting in lower rewards. DRL with a learning rate of $10^{-2}$ performs poorly because a high learning rate increases oscillations, the result uses sliding average of 200 times.
	
\begin{figure}[t]
		\centering
		\subfloat[The car speed $v=$ 40 km/h]{\includegraphics[width=0.495\columnwidth]{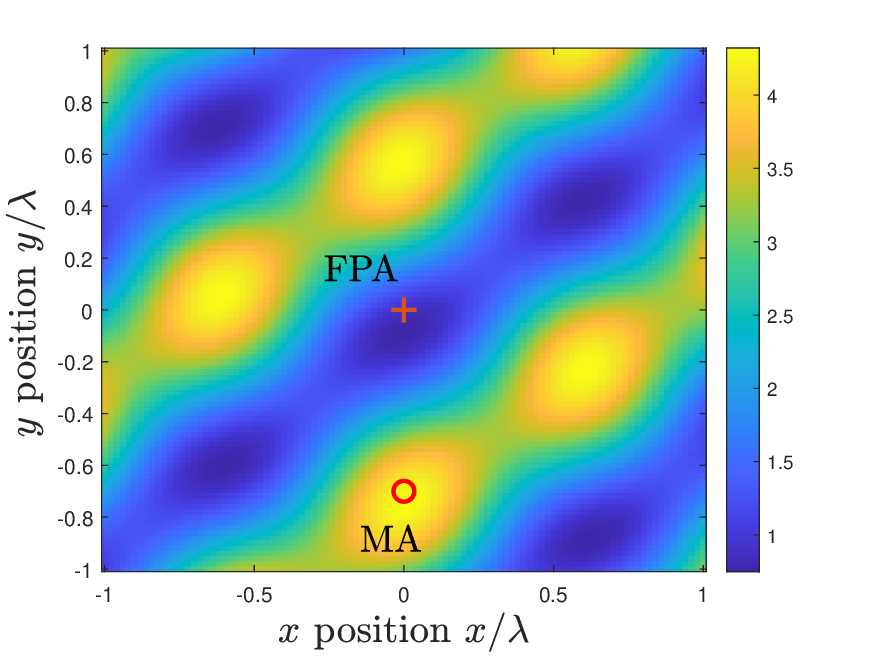}%
			\label{visi40}}
		\hfil
		\subfloat[The car speed $v=$ 70 km/h]{\includegraphics[width=0.495\columnwidth]{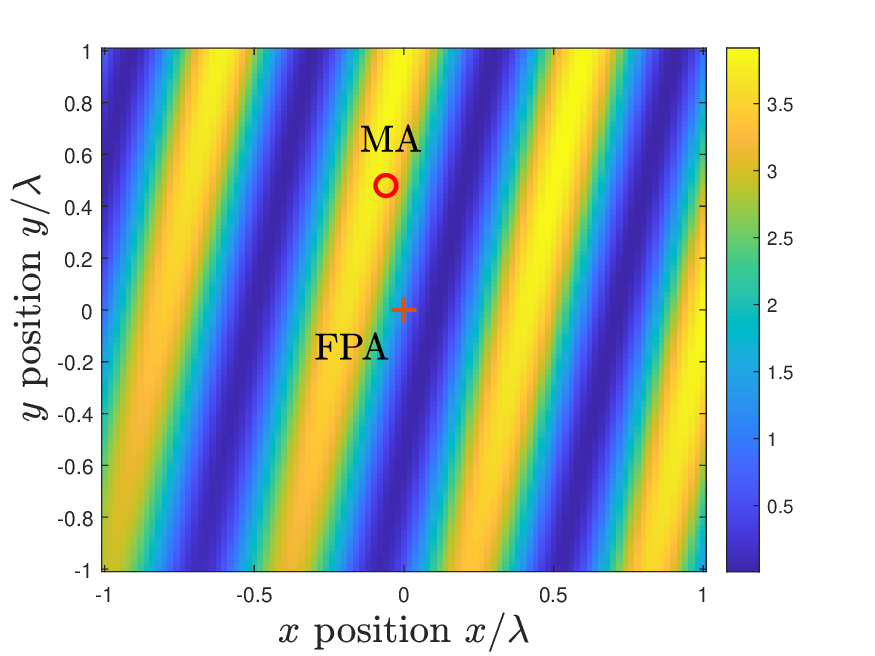}%
			\label{visi70}}
		\hfil
		\caption{Channel gain heatmap with MA and FPA positions in two different channel environments for $\alpha = 10^{-7}$.}
		\label{heatmapfig}
		\vspace{-0.3cm}
	\end{figure}

	As shown in Fig.~\ref{heatmapfig}, we plot the channel gain figure with MA and FPA positions for a learning rate of $10^{-7}$ in different environments. We can see that antenna mobility enables the MA to move away from deep fading regions and approach locations with stronger constructive combining. Whereas the FPA position is fixed, therefore suffers from severe attenuation.
	
	\begin{figure}[t]
		\centering
		\subfloat[The car speed $v=$ 40 km/h]{\includegraphics[width=0.495\columnwidth]{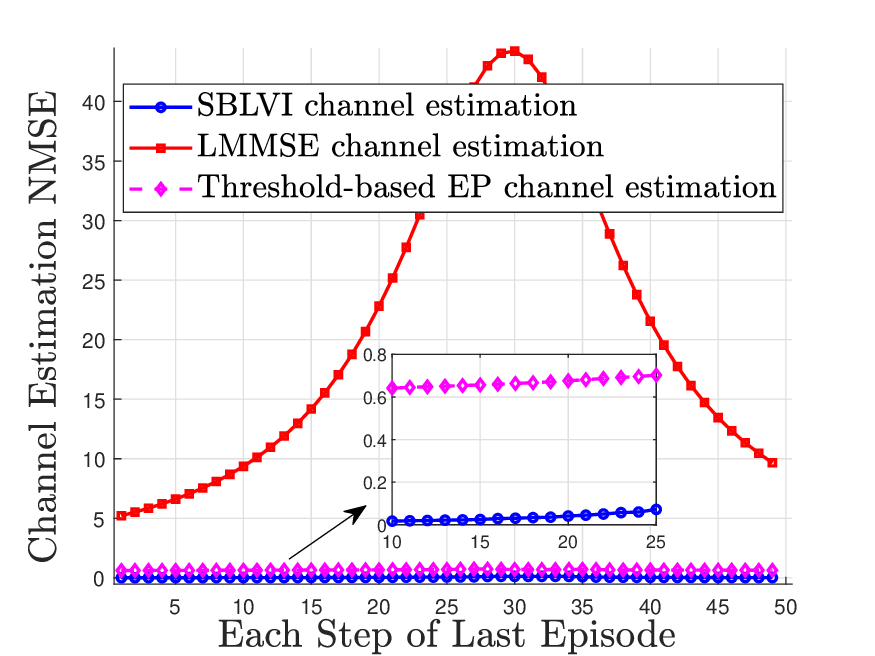}%
			\label{est_40kmph}}
		\hfil
		\subfloat[The car speed $v=$ 144 km/h]{\includegraphics[width=0.495\columnwidth]{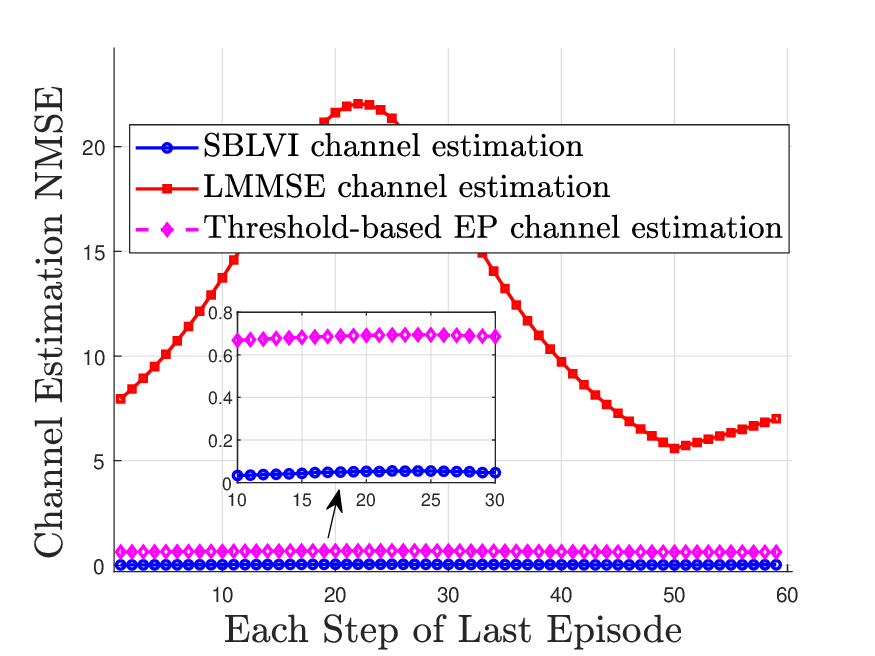}%
			\label{est_144kmph}}
		\hfil
		\caption{NMSE comparison in two different environments for $\alpha = 10^{-7}$.}
		\label{channelestcomp}
		\vspace{-0.3cm}
	\end{figure}
	
	We compare the different channel estimation methods under different environments using NMSE as performance metric in Fig.~\ref{channelestcomp}. We can see that the SBLVI method proposed in our design has obvious advantages over the traditional LMMSE and threshold-based EP methods in dynamic environment.
		
	\section{Conclusions}
	In this work, we studied an MA-assisted OTFS system under imperfect CSI. By applying MA position optimization, the proposed system mitigates deep fading and achieves higher channel gain performance than FPA. We proposed the SBLVI method for accurate channel estimation, and a DRL algorithm was used for dynamic MA position optimization. Simulation results showed that SBLVI significantly improves estimation accuracy over the benchmark methods, DRL-based MA position optimization achieves substantial improvements in channel gain. Our design can enable high-performance wireless communication with MA in future dynamic scenarios.
	
\bibliographystyle{IEEEtran}
\bibliography{refs}

\end{document}